\documentclass[%
reprint,
amsmath,amssymb,
pra,
]{revtex4-2}

\usepackage{graphicx}
\usepackage{dcolumn}
\usepackage{bm}
\usepackage[T1]{fontenc}
\usepackage{lineno}
\usepackage{xcolor}

\usepackage{multirow}
\usepackage{graphicx}
\usepackage{braket}
\usepackage{verbatim}
\usepackage{booktabs}
\usepackage{hyperref}
\usepackage{subfiles}
\newlength{\myspacing}
\begin{document}

\preprint{APS/123-QED}

\title{Simultaneous sub-Doppler laser cooling and optical trapping of bosonic $^{39}$K-$^{133}$Cs and $^{41}$K-$^{133}$Cs mixtures}

\author{Mateusz Boche{\'n}ski}
\altaffiliation[Present address: ]{Department of Physics, Durham University, South Road, Durham DH1 3LE, United Kingdom}
\affiliation{Institute\ of\ Experimental\ Physics,\ University\ of\ Warsaw,\ Pasteura\ 5,\ 02-093\ Warsaw,\ Poland}

\author{Jakub Dobosz}
\altaffiliation[Present address: ]{Department of Physics, Stockholm University, 10691 Stockholm, Sweden}
\affiliation{Institute\ of\ Experimental\ Physics,\ University\ of\ Warsaw,\ Pasteura\ 5,\ 02-093\ Warsaw,\ Poland}

\author{Koray Din\c{c}er}
\altaffiliation[Present address: ]{eleQtron GmbH, Siegen, Germany}
\affiliation{Institute\ of\ Experimental\ Physics,\ University\ of\ Warsaw,\ Pasteura\ 5,\ 02-093\ Warsaw,\ Poland}

\author{Paweł Arciszewski}
\altaffiliation[Present address: ]{Institute of Electronic Systems, Warsaw University of Technology, 00-665 Warsaw, Poland}
\affiliation{Institute\ of\ Experimental\ Physics,\ University\ of\ Warsaw,\ Pasteura\ 5,\ 02-093\ Warsaw,\ Poland}

\author{Mariusz Semczuk}%
\email{msemczuk@fuw.edu.pl}
\altaffiliation[Also at: ]{Institute of Electronic Systems, Warsaw University of Technology, 00-665 Warsaw, Poland}
\affiliation{Institute\ of\ Experimental\ Physics,\ University\ of\ Warsaw,\ Pasteura\ 5,\ 02-093\ Warsaw,\ Poland}

\date{\today}

\begin{abstract}

We report simultaneous sub-Doppler cooling and optical dipole trapping of $^{39}$K-Cs and $^{41}$K-Cs mixtures. Both mixtures are cooled to temperatures $\sim10$~\textmu K, achieving performance comparable to that obtained with each species individually. To the best of our knowledge, this constitutes the first realization of a laser-cooled and optically trapped $^{41}$K-Cs mixture. For the $^{39}$K-Cs mixture, we additionally implement parallel spin-resolved Feshbach spectroscopy enabled by Stern--Gerlach separation. 

 The measurements were performed in a single-chamber setup, with atoms loaded directly from the background gas. For potassium isotopes, we implement sub-Doppler cooling with $D_1$-line gray molasses, while for cesium the entire cooling sequence is implemented using $D_2$ transitions. We spin-polarize the atoms and confine Bose-Bose mixtures in a 1064~nm optical dipole trap.

Using the resulting $^{39}$K-Cs samples, we observe 14 heteronuclear Feshbach loss features. Five agree with resonances reported previously, while nine have, to the best of our knowledge, not been observed experimentally before, including $p$-wave features and resonances in additional spin channels.

The shorter lifetime of the $^{41}$K-Cs mixture in the optical dipole trap currently hinders systematic Feshbach spectroscopy, which we therefore do not pursue in this work. To characterize this limitation quantitatively, we study the decay dynamics of $^{39}$K-Cs and $^{41}$K-Cs mixtures under comparable temperature and density conditions, revealing a substantially stronger nonexponential loss in the $^{41}$K-Cs mixture.

The demonstrated preparation of ultracold $^{41}$K-Cs mixtures provides a starting point for Feshbach-resonance and photoassociation spectroscopy of this previously unexplored isotopologue. Such measurements are essential for identifying suitable pathways for magnetoassociation and coherent optical transfer, and ultimately for the production of ultracold ground-state $^{41}$KCs molecules.
 
\end{abstract}

\maketitle

\section{Introduction}

Ultracold mixtures of distinct atomic species provide access to interaction regimes and molecular states unavailable in single-species systems. Control of interspecies interactions through magnetic Feshbach resonances has enabled sympathetic cooling, studies of strongly interacting mixtures, and the production of ultracold heteronuclear molecules~\cite{modugno2001bose, hadzibabic2002two, aubin2006rapid, roati2002fermi, jochim2003bose, greiner2003emergence}. In particular, the preparation of polar ground-state molecules by magnetoassociation followed by coherent optical transfer relies on the ability to produce dense, low-temperature atomic mixtures and on accurate knowledge of their interspecies scattering properties~\cite{kuznetsova2009efficient, ni2008high}. Such molecules provide long-range and anisotropic dipole--dipole interactions that are attractive for studies of dipolar many-body physics, quantum simulation, and controlled ultracold chemistry~\cite{gadway2016strongly, moses2017new, cornish2024quantum}.

Among alkali-metal mixtures, potassium--cesium systems are particularly attractive because the three stable potassium isotopes provide access to both bosonic and fermionic KCs molecules within the same general experimental platform. Ground-state KCs molecules are predicted to possess a permanent electric dipole moment of 1.92~D, placing them among the most strongly dipolar alkali dimers~\cite{aymar2005calculation}. The different potassium isotopes also provide substantially different collisional properties while retaining nearly identical optical requirements.

Experimental progress, however, has so far been concentrated predominantly on $^{39}$K-Cs. Feshbach spectroscopy of this mixture~\cite{PhysRevA.95.022715,beulenkamp2025formation} has recently been followed by the production of ground-state $^{39}$KCs molecules~\cite{zamarski2025spectroscopy}. To the best of our knowledge, a laser-cooled $^{41}$K-Cs mixture has not previously been realized in either a magneto-optical trap or an optical dipole trap. Establishing efficient preparation of this isotopologue extends the experimentally accessible KCs systems beyond the isotope combination studied to date.

In this work, we demonstrate simultaneous sub-Doppler cooling and optical dipole trapping of both $^{39}$K-Cs and $^{41}$K-Cs mixtures in a single-chamber apparatus, with atoms loaded directly from background vapor. Potassium is cooled using $D_1$ gray molasses, while the cesium cooling sequence uses only light near the $D_2$ line. More than $10^6$ atoms per species are obtained in the optical dipole trap at temperatures below $15~\mu$K. This constitutes the first realization of an ultracold $^{41}$K-Cs mixture in an optical dipole trap.

We use the resulting samples to perform multichannel Feshbach spectroscopy of $^{39}$K-Cs. A spin-resolved detection method based on Stern--Gerlach separation allows the three $F=1$ potassium spin components to be measured independently after the same interaction cycle, providing the loss spectra of three entrance channels simultaneously. Using this approach together with conventional spectroscopy of spin polarized samples, we observe 14 heteronuclear loss features, five corresponding to previously reported resonances and nine, to the best of our knowledge, not previously observed experimentally. The additional features include $p$-wave resonances and resonances in spin channels not covered by earlier measurements.

For $^{41}$K-Cs, we observe substantially faster decay in the optical dipole trap than for $^{39}$K-Cs under comparable conditions. The nonexponential decay is consistent with an additional density-dependent loss contribution and presently limits systematic Feshbach spectroscopy of this mixture. We therefore compare the decay dynamics of the two isotopologues, establishing the experimental conditions that must be improved before a comparable resonance search can be performed for $^{41}$K-Cs.

Together, these results extend ultracold K-Cs experiments beyond $^{39}$K-Cs, provide additional constraints on the $^{39}$K-Cs interaction potentials, and establish a route toward systematic interaction spectroscopy and molecule formation with multiple KCs isotopologues.

\section{Experimental setup and single species cooling\label{sec:exp_set}}

The experiments are performed in a single-chamber apparatus, with potassium and cesium atoms loaded directly from background vapor using atomic dispensers located near the magneto-optical trap. The uncoated glass cell is vertically oriented and pumped with an ion pump and a titanium sublimation pump. Magnetic fields are generated by five pairs of coils. Three pairs are used to compensate for residual ambient magnetic fields and to define a quantization axis. The remaining two pairs operate in Helmholtz and anti-Helmholtz configurations and generate, respectively, the uniform magnetic field used for Feshbach spectroscopy and the magnetic-field gradient required during stages such as magneto-optical trap loading. The uniform magnetic field that can safely be generated by the coils without additional water cooling is around 500~G, which is sufficient to reach many Feshbach resonances in both $^{39}$K-Cs and $^{41}$K-Cs mixtures and fermionic $^{40}$K-Cs~\cite{PhysRevA.90.032716,PhysRevA.95.022715, beulenkamp2025formation}.

Unless stated otherwise, all detunings are expressed in units of the natural linewidths of cesium $\Gamma_{\mathrm{Cs}} = 2\pi \times 5.223$~MHz, and potassium $\Gamma_{\mathrm{K}} = 2\pi \times 6.035$~MHz $D_2$-lines. Laser intensities are reported in units of $I_{\mathrm{s}}^{\mathrm{Cs}} = 1.65$~mW/$\mathrm{cm^2}$ for cesium, and $I_{\mathrm{s}}^{\mathrm{K}} = 1.75$~mW/$\mathrm{cm^2}$ for potassium~\cite{steck2003cesium, tiecke2010properties}. 

In Section~\ref{sec:exp_set} we limit our discussion only to essential information that directly ties to the work with the studied mixtures. Further details of the vacuum system, magnetic-field coils, parameters of the individual cooling stages and implemented technical solutions are summarized in the Supplemental Material~\cite{supp}.

\subsection{Laser cooling of potassium atoms}

The cooling procedures for the potassium isotopes and the laser system used have been discussed in detail in our previous work~\cite{bochenski2024magnetic}. Briefly, potassium atoms are first trapped in a magneto-optical trap using light close to the $D_2$-line frequency. After the MOT loading stage, the cloud is compressed by increasing the magnetic-field gradient and red-detuning the cooling and repumping beams. During the final part of the compression, $D_1$-line cooling and repumping beams are introduced, producing a D2D1-CMOT similar to that described in Refs.~\cite{salomon2014gray, chen2016production}. The $D_2$-line light is subsequently switched off and the atoms are cooled using $\Lambda$-enhanced gray optical molasses on the $D_1$ line. This stage allows for almost lossless temperature reduction to 8--13~\textmu K, depending on the isotope and technical aspects such as the alignment of the beams or drifts in power and polarization.

During gray optical molasses cooling, we switch on the optical dipole trap, formed by a single 1064~nm beam retro-routed so that it passes twice through the cloud of atoms and forms a crossed dipole trap with an intersection angle of 22.5$\mathrm{^o}$. The polarization of the beam is rotated by 90$\mathrm{^o}$ before its second pass through the cloud to prevent the formation of a lattice. The trap depth determined from parametric-heating measurements corresponds to approximately $U = k_{\mathrm{B}}\times87$~$\mu$K for potassium and $U = k_{\mathrm{B}}\times165$~$\mu$K for cesium.

Cooling in gray optical molasses, despite the favorable density, temperature, and atom number, yielded a transfer efficiency into the optical dipole trap of no more than 0.3\% for either bosonic potassium isotope. We therefore introduced an additional gray-molasses stage optimized specifically for transfer into the optical dipole trap. For clarity, we refer to this stage throughout the manuscript as Transfer-Optimized Gray Optical Molasses (TO-GMC), using this term only to distinguish the optimization goal from that of the preceding gray-molasses stage. During TO-GMC, the beam frequencies remain unchanged, while their intensities and the duration of the cooling stage are optimized for transfer into the optical dipole trap.

\begin{figure}[ht]
\centering
\includegraphics[width=\columnwidth]{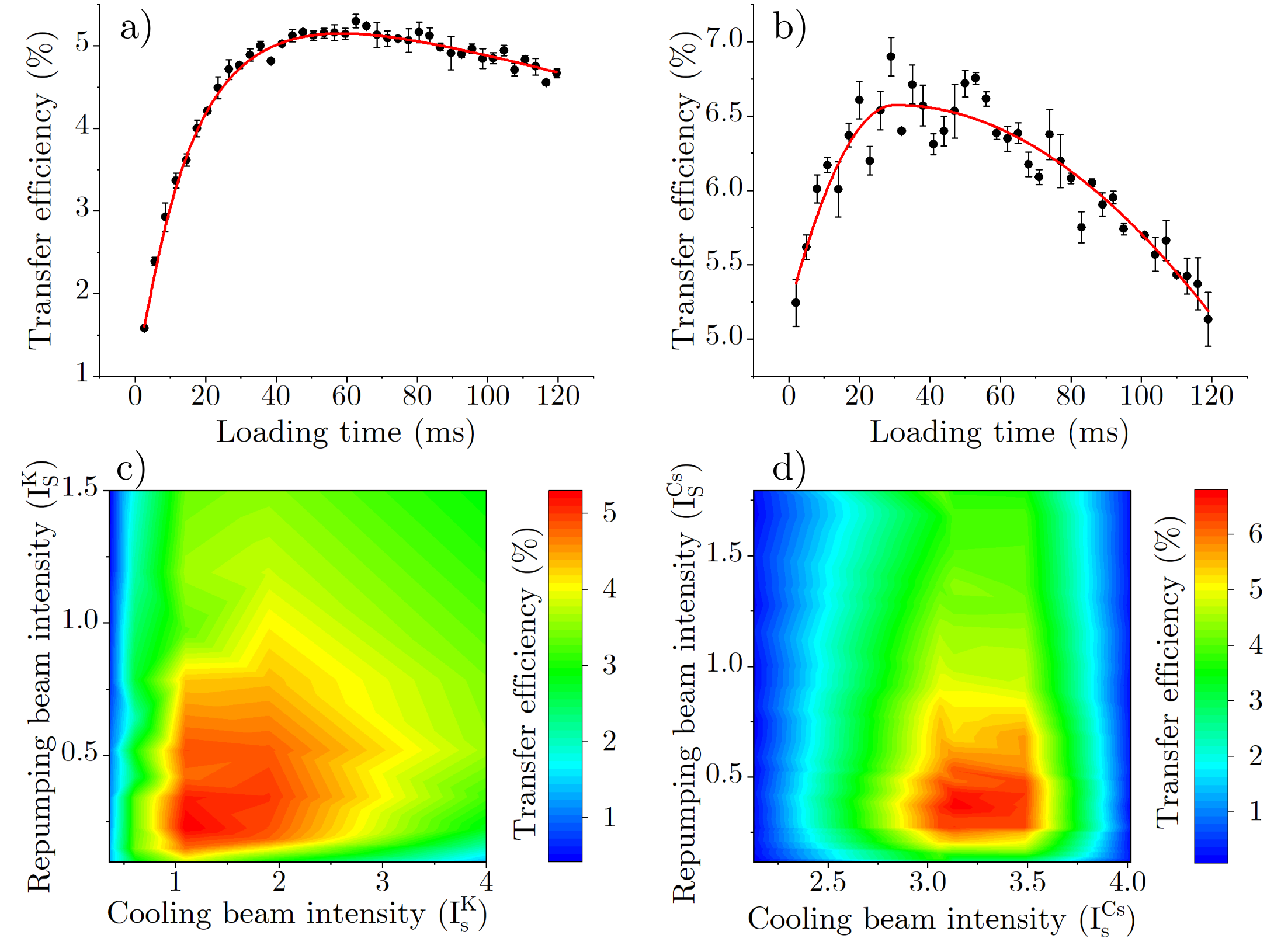}
\caption{Transfer efficiency from gray optical molasses into the optical dipole trap for $^{41}$K (a,c) and Cs (b,d) atoms. The efficiency is presented as a function of the loading time (a,b) and the intensities of gray molasses light (c,d). The red curve on the top plots is presented as a guide to the eye.}
\label{loading_K_Cs}
\end{figure}

The transfer efficiency as a function of the TO-GMC duration and beam intensities is shown in Fig.~\ref{loading_K_Cs}. For both bosonic isotopes, this additional stage substantially improves the loading efficiency, reaching approximately 5.3\% for $^{41}$K and 1.8\% for $^{39}$K relative to the number of atoms at the end of gray-optical-molasses cooling. In single-species operation, we finally obtain approximately $2.0\times10^6$ $^{41}$K atoms and $3.6\times10^6$ $^{39}$K atoms in the optical dipole trap, at temperatures of approximately 7.5~$\mu$K.

The achieved transfer efficiency is promising, particularly given the low initial temperature and simplicity of the method. For $^{39}$K, gray molasses followed by magnetic trapping has enabled transfer efficiencies of 2.7--8\% into an optical dipole trap, with initial temperatures ranging from 48~\textmu K to almost 1~mK~\cite{grobner2016new,Landini_direct}. Degenerate Raman sideband cooling has achieved efficiencies of 12\% for fermionic potassium~\cite{zohar2022degenerate}, but requires additional optical elements and beam paths. In comparison, our method yields temperatures below 15~\textmu K and uses the same optical setup as the MOT beams.

The TO-GMC approach is also well suited to multi-species experiments. Its implementation for one species requires only low-intensity light far detuned from the main absorption lines of the other species and does not introduce additional optical paths. Moreover, gray molasses cooling has been demonstrated for a broad range of atomic species, including $^{6}$Li, $^{7}$Li, $^{87}$Rb, and $^{23}$Na~\cite{burchianti2014efficient,grier2013lambda,rosi2018lambda,colzi2016sub}, suggesting that similar transfer schemes could be applicable beyond potassium.

The last stage of potassium sample preparation is spin polarization in the selected magnetic sub-level. Depending on the direction of the quantization axis, we obtain a sample containing approximately 80\% of atoms in the $^2S_{1/2}F=1,~m_f=1$ or $m_f=-1$ state, with most of the remaining atoms occupying the $^2S_{1/2}F=1,~m_f=0$ state. Details of the optical-pumping procedure are given in the Supplemental Material.

\subsection{Laser cooling of cesium atoms}

For cesium, all laser-cooling stages use transitions within the $D_2$ line. The atoms are first trapped in a magneto-optical trap and subsequently compressed by increasing the magnetic-field gradient and changing the cooling and repumping detunings. This is followed by gray-molasses cooling using a cooling beam blue-detuned from the $F=4\to F'=4$ transition and a repumping beam close to the $F=3\to F'=3$ transition. After 8~ms of gray-molasses cooling, we obtain approximately $7.5\times10^7$ atoms at a temperature of $23$~\textmu K.

As in the potassium case, we introduce a TO-GMC stage to improve transfer into the optical dipole trap. The beam frequencies are kept unchanged and their intensities and the cooling time are optimized for loading. The measured transfer efficiency as a function of these parameters is shown in Fig.~\ref{loading_K_Cs}. Using a transfer time of $t = 40$~ms, we obtain approximately $5.1\times10^6$ cesium atoms in the optical dipole trap at a temperature of approximately 12~$\mu$K.

Finally, cesium atoms are spin-polarized in the energetically lowest magnetic sublevel. We obtain samples containing approximately 90\% of atoms in the $F=3, m_{\mathrm{F}}=3$ state, with most of the remaining atoms occupying the $F=3, m_{\mathrm{F}}=2$ state. 

\section{Simultaneous cooling and collisional studies of K-Cs mixtures}

In our multi-species experimental system, the optimal cooling strategy for the potassium-cesium mixtures depends strongly on the abundance of the potassium isotope in the source. For the less abundant $^{41}$K isotope, the cooling sequence is optimized to maximize the number of potassium atoms in the optical dipole trap, even at the cost of a modest reduction in the number of cesium atoms. In contrast, in the $^{39}\mathrm{K}$–Cs mixture, both species are captured in comparable numbers in the magneto-optical trap, allowing for a balanced optimization that yields similar atom numbers in the dipole trap.

The mixtures of $^{39}$K-Cs and $^{41}$K-Cs are prepared using the same general procedures as those used to prepare single-species samples (see Sec.~\ref{sec:exp_set} and Supplementary Material~\cite{supp}), with parameters optimized for simultaneous cooling. We observe a slight reduction in the number of trapped atoms when both species are present, which may arise from interspecies collisions. As noted in the previous section, the optimal compression of potassium and cesium occurs at different magnetic-field gradients. We therefore choose the CMOT parameters to prioritize the density of the more dilute potassium cloud while retaining a sufficiently large cesium sample for subsequent loading into the optical dipole trap.

Finally, we obtain both K-Cs mixtures in the optical dipole trap at temperatures typically in the range of 10-15\textmu K (as determined by the time-of-flight method), comparable to those obtained in single-species operation. The $^{39}$K-Cs mixture contains $N_{\mathrm{39K}}=3.1\times10^6$ and $N_{\mathrm{Cs}}=3.8\times10^6$ atoms, while for $^{41}$K-Cs we obtain $N_{\mathrm{41K}}=1.1\times10^6$ and $N_{\mathrm{Cs}}=4.3\times10^6$ atoms.

\subsection{Feshbach spectroscopy of the $^{39}$K-Cs mixture}
For clarity, in the following part of this paper the magnetic sublevels of the hyperfine ground states are labeled alphabetically in order of increasing energy. Specifically, for $^{39}$K and $^{41}$K atoms we use $\ket{a}$, $\ket{b}$, and $\ket{c}$ to denote the low-field states $^2S_{1/2},F=1,m_{\mathrm{F}}=1,0,-1$, respectively. Similarly, for $^{133}$Cs the labels $\ket{a}$, $\ket{b}$, $\ldots$ denote the states $^2S_{1/2},F=3,m_{\mathrm{F}}=3,2,\ldots$, respectively. Throughout this section, the potassium state is specified first in the notation for a collision channel.

Feshbach spectroscopy is performed using two different approaches. First, a spin-polarized mixture $^{39}\mathrm{K}\ket{a}+\mathrm{Cs}\ket{a}$ (or $^{39}\mathrm{K}\ket{c}+\mathrm{Cs}\ket{a}$) is loaded into the optical dipole trap. Then, over ~100~ms, a magnetic field is turned on and held for 500~ms. After that time, we switch off the magnetic field in 100~ms, and fluorescence imaging is performed, providing information on the number of remaining atoms. Although both species are imaged, the resonance features are substantially more pronounced in the number of potassium atoms, which we use to identify and characterize the resonances.

The standard approach is limited by incomplete spin polarization. In our system, a small fraction of both potassium and cesium atoms remains in undesired magnetic sublevels after optical pumping (about 10\% for cesium). Consequently, a nominally single-channel measurement may contain weak loss features originating from additional entrance channels. State purification can be improved using, for example, magnetic trapping or RF transfer~\cite{PhysRevA.95.022715, cordova2018quantum}, but these approaches require additional preparation steps.

To distinguish resonances involving Cs in the $\ket{a}$ and $\ket{b}$ states, we additionally prepare a cesium sample in which the population of the $\ket{a}$ state is strongly suppressed. To do so, first, the cesium atoms are polarized in the $F = 3$,~$m_F=3$ state. Then, a magnetic field of $B = 88$~G is applied. In this field, we introduce a 1~ms pulse of $\pi$-polarized light with an intensity of $0.6I_s^{\text{Cs}}$. The light is blue-detuned by $9\Gamma_{\text{Cs}}$ from the zero-field $F=3 \to F'=3$ transition. At this detuning, the beam becomes resonant with the high-field transition $F=3,~m_F=3 \to F'=3,~m_F'=3$, while the closest allowed transition, $F=3,~m_F=2 \to F'=3,~m_F'=2$, remains detuned by approximately $3\Gamma_{\text{Cs}}$.

To prevent cesium atoms from accumulating in the $F=4$ state during the pumping process, a MOT cooling light is applied simultaneously. Finally, atoms remaining in the $F=3,m_F=3$ state are removed, resulting in a sample predominantly occupying the $F=3,~m_F=2$ state. Comparison of spectra obtained with the two cesium preparations allows us to distinguish loss features associated with the Cs $\ket{a}$ and $\ket{b}$ entrance channels.

For potassium, instead of preparing a separate sample in the $F=1,m_F=0$ state for each collision channel, we implement spin-resolved detection based on Stern--Gerlach separation. A mixture of potassium spin states together with a spin-polarized cesium sample is loaded into the optical dipole trap, with the potassium population distributed approximately equally among the $F=1,m_F=\{1,0,-1\}$ states.

\begin{figure}[ht]
\centering
\includegraphics[width=\linewidth]{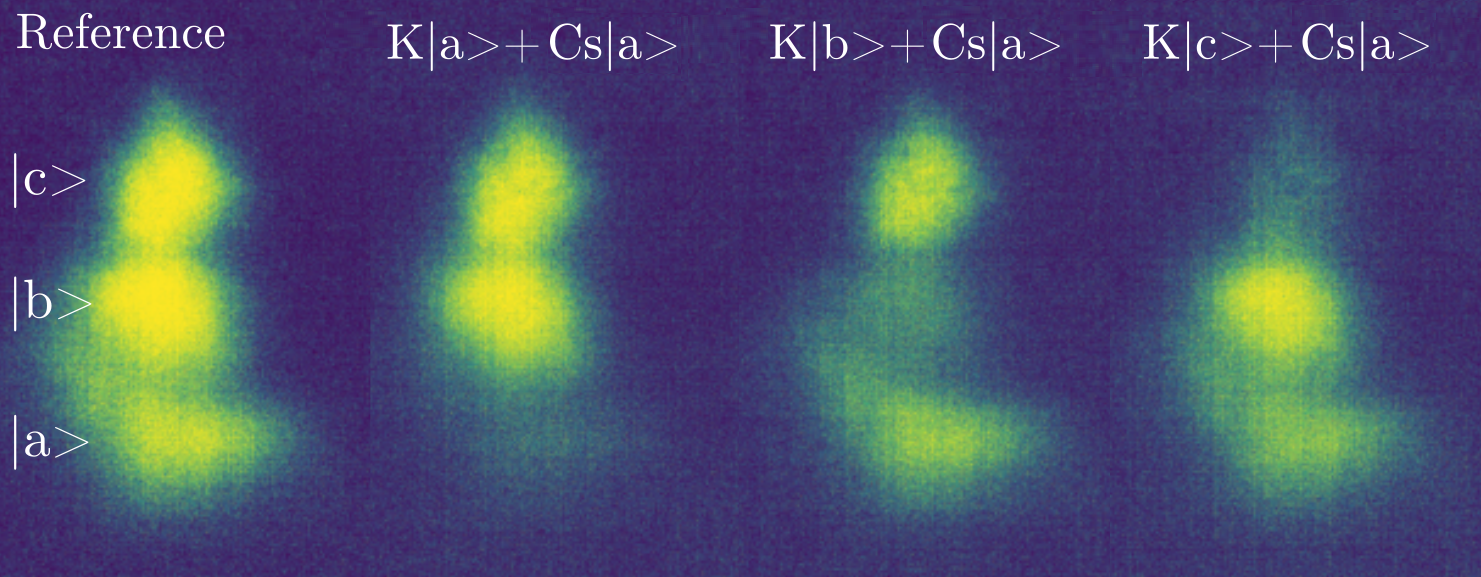}
\caption{Fluorescence images of magnetically separated $^{39}$K samples after heteronuclear Feshbach spectroscopy. Each image shows three spatially separated potassium clouds corresponding to the magnetic sublevels of the $ F=1 $ state with $ m_F = \{1, 0, -1\} $, denoted as $\ket{a}$, $\ket{b}$ and $\ket{c}$, respectively. From left to right: reference image (no loss feature), and three cases showing atom loss due to Feshbach resonances between ${}^{39}\mathrm{K}$ and ${}^{133}\mathrm{Cs}$ in the channels indicated at the top of each panel. }
\label{separation}
\end{figure}

The Feshbach-spectroscopy sequence then follows the procedure described above. However, before imaging, the potassium spin components are spatially separated. A homogeneous magnetic field of 17~G is applied over 10~ms together with a magnetic field gradient of 36~G/cm ramped up over 6~ms. After release from the optical dipole trap, an 8~ms separation time provides sufficient spatial resolution of the three spin components while avoiding excessive cloud expansion. The atom number in each component is then determined independently. An example of spatially separated atomic clouds in the vicinity of a Feshbach resonance and far from a resonance is shown in Fig.~\ref{separation}.

\begin{figure}[ht]
\centering
\includegraphics[width=\linewidth]{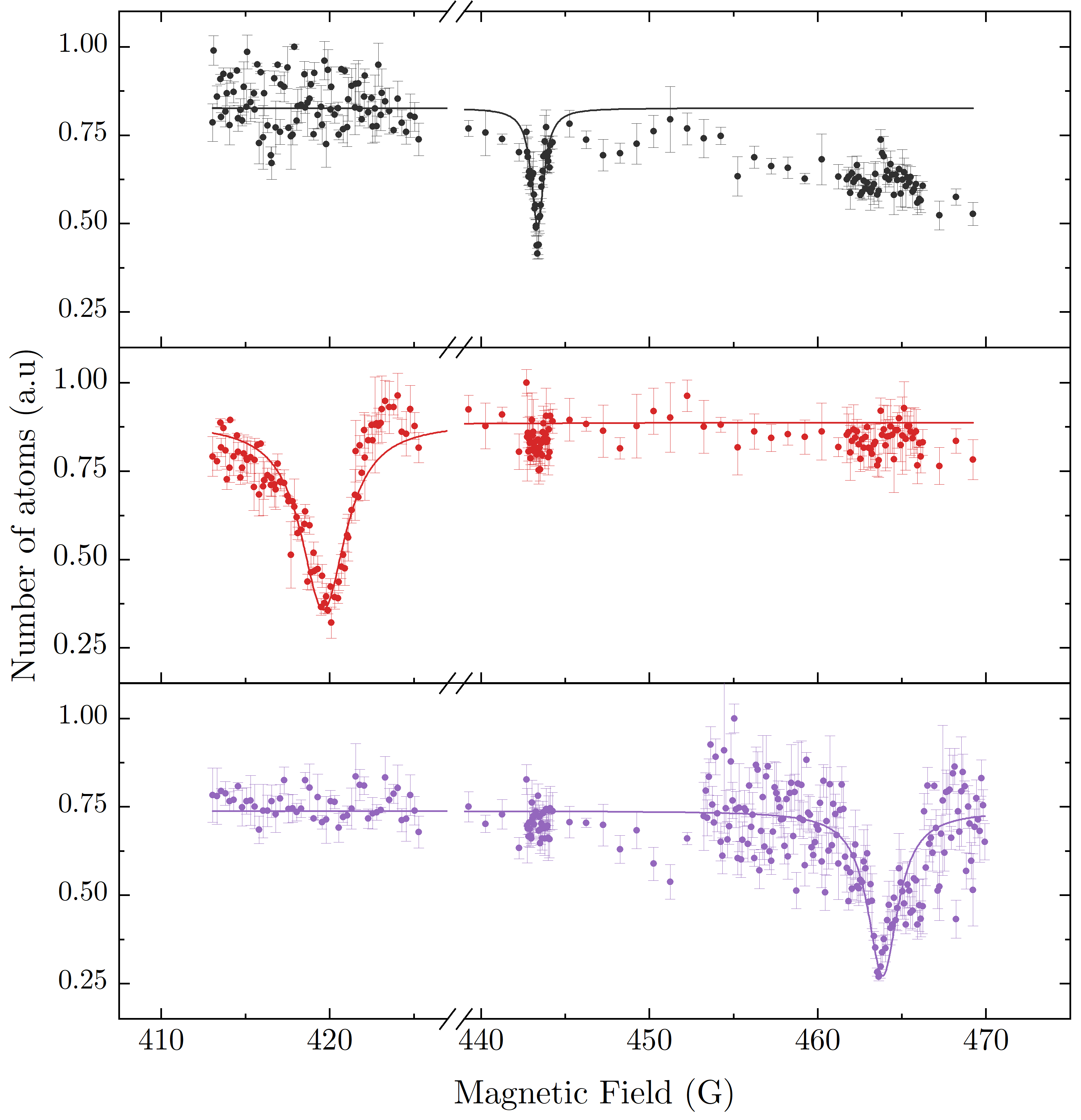}
\caption{Number of atoms as a function of magnetic field for three collision channels $\ket{a}+\ket{a}$ (top, black), $\ket{b}+\ket{a}$ (middle, red), and $\ket{c}+\ket{a}$ (bottom, purple) present simultaneously in the trapped sample. The first state label denotes the magnetic sublevel of $^{39}$K and the second that of $^{133}$Cs. The measurements illustrate the simultaneous spin-resolved detection of Feshbach loss features using Stern-Gerlach separation.}
\label{separation_FSH_plot}
\end{figure}

Fig.~\ref{separation_FSH_plot} shows representative loss spectra measured simultaneously for the three spatially separated spin components, corresponding to the collision channels $\ket{a}+\ket{a}$, $\ket{b}+\ket{a}$, and $\ket{c}+\ket{a}$. Each loss feature is confined to its corresponding spin channel, with no discernible feature at the same magnetic field in the other components. The method therefore provides the loss spectra of three potassium spin channels within the same experimental sequence, reducing by up to a factor of three the number of experimental cycles required for a multichannel scan. It also eliminates the need for challenging optical pumping into the $F=1,m_F=0$ state and avoids systematic differences that may arise when individual channels are measured in separate scans.  

\begin{figure*}[ht]
\centering
\includegraphics[width=\linewidth]{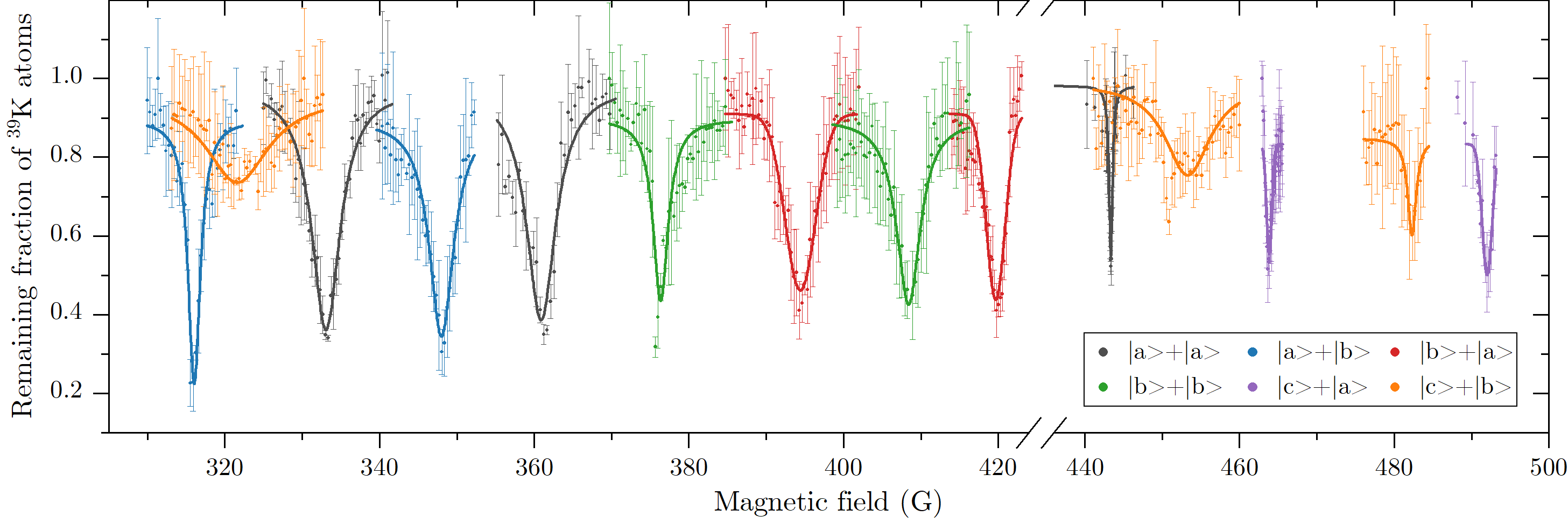}
\caption{Overview of losses in $^{39}$K atoms present due to the heteronuclear $^{39}$K-Cs Feshbach resonances. Empirical Lorentzian profiles are fitted to the loss spectra to determine their centers and FWHM. Different colors correspond to different collision channels, as presented in the legend. Homonuclear losses are omitted for clarity.}
\label{39KCs_FSH_total}
\end{figure*}

With the methods described above, we observe 14~loss features assigned to heteronuclear Feshbach resonances, as shown in Fig.~\ref{39KCs_FSH_total}. Empirical Lorentzian fits provide the loss-feature positions and FWHM, which are listed in Table~\ref{tab:FSH_table_summary} together with theoretical resonance positions and partial-wave assignments. The uncertainties quoted for the experimental resonance positions correspond to the statistical uncertainties of the fitted line center. The uncertainty of the RF-calibrated magnetic field is below 50~mG and is therefore negligible compared with the fit uncertainties. Five of the observed features correspond to resonances reported previously, while nine have, to the best of our knowledge, not been observed experimentally before.

To obtain theoretical resonance positions, we performed coupled-channel calculations using the MOLSCAT package at a collision energy of $E_{\mathrm{coll}}/k_{\mathrm{B}}=100$~nK~\cite{hutson2019molscat}. The interaction potential used in the calculations, referred to as B2025, is based on the results presented in~\cite{ferber2013long}, with parameters refined in~\cite{beulenkamp2025formation}. For 11 of the 14 observed features, the measured loss positions agree with the coupled-channel calculations to within 0.7~G. The remaining three features show larger deviations of 2.9-3.3~G.

These discrepancies substantially exceed the uncertainty of our magnetic-field calibration, which we determine from RF spectroscopy of magnetic-dipole transitions in $^{41}$K to scale as approximately 0.033~G per 100~G of applied field. In addition, two of the three discrepant features are assigned to $p$-wave resonances and may therefore be more sensitive to finite-temperature effects. A quantitative interpretation of the observed deviations will require further investigation, including finite-temperature calculations.

Particular attention is required for the resonance observed near 316~G. A resonance in this region was previously reported in the $\ket{b}+\ket{b}$ channel~\cite{beulenkamp2025formation}. In our measurements, however, a loss feature at 315.95(15)~G is observed only for the $\ket{a}+\ket{b}$ mixture, while no corresponding feature is detected in the $\ket{b}+\ket{b}$ channel. Our coupled-channel calculations indicate that two distinct resonances are expected in this region: an $L=1$ resonance in the $\ket{a}+\ket{b}$ channel at 315.54~G, consistent with our observation, and a separate resonance in the $\ket{b}+\ket{b}$ channel near 315.77~G, consistent with the previously reported resonance. This case illustrates the utility of spin-resolved detection: the two predicted resonances are very close in magnetic field, but occur in different entrance channels and can therefore be distinguished directly through their spin dependence.

\begin{table}[ht]
\begin{tabular}{cccccccc}
                            &                           & Theory    &                 \multicolumn{2}{c}{Experiment}                  & \multicolumn{2}{c}{Experiment} \\
                           &                            & B2025        & \multicolumn{2}{c}{This work} & \multicolumn{2}{c}{Ref.~\cite{PhysRevA.95.022715, beulenkamp2025formation}}      &            \\ \midrule \midrule
K                 & Cs                & $\mathrm{B_{res}}$~(G) & $\mathrm{B_{res}}$~(G) & $\Delta_{\mathrm{loss}}$~(G) & $\mathrm{B_{res}}$~(G) & $\Delta_{\mathrm{loss}}$~(G) & L \vspace{0.1 cm} \\  \midrule
\multirow{3}{*}{$\ket{a}$} & \multirow{3}{*}{$\ket{a}$} & 361.66                & 360.95(18)          & 2.67(38)           & 361.66(1)              & 3.2(4)             & 0          \\
                           &                            & 443.45                & 443.28(15)           & 0.38(6)            & 442.6(3)             & 0.28(3)            & 0          \\
                           &                            & 332.94                & 333.32(12)             & 1.55(40)           & -                     & -                  & 1         \vspace{\myspacing} \\ 
\multirow{2}{*}{$\ket{a}$} & \multirow{2}{*}{$\ket{b}$} & 348.12                & 347.83(18)            & 3.96(42)           & -                     & -                  & 0          \\
                           &                            & 315.54                & 315.95(15)            & 1.60(31)           & -            & -                  & 1        \vspace{\myspacing}  \\ 
\multirow{2}{*}{$\ket{b}$} & \multirow{2}{*}{$\ket{a}$} & 420.24                & 419.70(15)          & 2.43(16)           & 419.3(3)              & 3.0(5)             & 0          \\
                           &                            & 391.15                & 394.42(17)            & 3.94(25)           & -                     & -                  & 1        \vspace{\myspacing}  \\ 
\multirow{2}{*}{$\ket{b}$} & \multirow{2}{*}{$\ket{b}$} & 408.78                & 408.40(18)            & 3.70(43)           & -                     & -                  & 0          \\
                           &                            & 375.75                & 376.36(16)            & 2.39(34)           & -                     & -                  & 1        \vspace{\myspacing}  \\ 
\multirow{2}{*}{$\ket{c}$} & \multirow{2}{*}{$\ket{a}$} & 492.24                & 491.94(18)            & 2.06(32)           & 491.5(1)              & 2.1(4)             & 0          \\
                           &                            & 463.35                & 463.99(16)            & 1.82(30)           & 463.251(5)            & -                  & 1        \vspace{\myspacing}  \\ 
\multirow{3}{*}{$\ket{c}$} & \multirow{3}{*}{$\ket{b}$} & 325.24                 & 322.32(38)            & 8.07(1.68)         & -                     & -                  & 0          \\
                           &                            & 482.62                 & 482.27(20)            & 2.16(45)           & -                     & -                  & 0          \\
                           &                            & 450.17                & 453.26(30)            & 8.84(1.16)         & -                     & -                  & 1          \\ \hline

\end{tabular}

\caption{Heteronuclear Feshbach resonances observed in the $^{39}$K–Cs mixture. The table lists the collision channels, theoretical resonance positions calculated using the B2025 interaction potentials \cite{beulenkamp2025formation}, experimental loss-feature positions ($B_{\mathrm{res}} $) and fitted full widths at half maximum ($\Delta_{\mathrm{loss}}$) extracted from empirical Lorentzian fits to the loss spectra shown in Fig.~\ref{39KCs_FSH_total}. Previously reported resonances from Ref.~\cite{PhysRevA.95.022715, beulenkamp2025formation} are also included for comparison. The final column indicates the orbital angular momentum $L$ associated with each resonance. The fitted $\Delta_{\mathrm{loss}}$ values characterize the widths of the measured atom-loss features and should not be identified with the magnetic widths $\Delta$ of the underlying Feshbach resonances.}
\label{tab:FSH_table_summary}
\end{table}

\subsection{\texorpdfstring{Lifetime of the $^{41}$K-Cs mixture}{41K-Cs mixture}}

Although the $^{41}$K-Cs mixture can be loaded efficiently into the optical dipole trap, we observe substantially faster atom loss than for $^{39}$K-Cs. We therefore compare the stability of the two isotopic mixtures under similar trapping and spin-polarization conditions.

For the lifetime measurements, we use samples with $N_{\mathrm{41K}}=1.0(1)\times10^6$ and $N_{\mathrm{Cs}}=1.0(1)\times10^6$ atoms for the $^{41}$K-Cs mixture, and $N_{\mathrm{39K}}=1.2(1)\times10^6$ and $N_{\mathrm{Cs}}=1.0(1)\times10^6$ atoms for the $^{39}$K-Cs mixture, corresponding to the peak densities of $n_{\mathrm{41K}}=6.1\times10^{11}$~atoms/cm$^3$, $n_{\mathrm{39K}}=7.4\times10^{11}$~atoms/cm$^3$, and $n_{\mathrm{Cs}}=1.4\times10^{11}$~atoms/cm$^3$. Both mixtures are predominantly prepared in the $^{39,41}\mathrm{K}\ket{a}+\mathrm{Cs}\ket{a}$ entrance channel and held in the optical dipole trap at a magnetic field of $B=5$~G. The atom-number decay is then measured as a function of trapping time.

\begin{figure}[ht] \centering 
\includegraphics[width=\columnwidth]{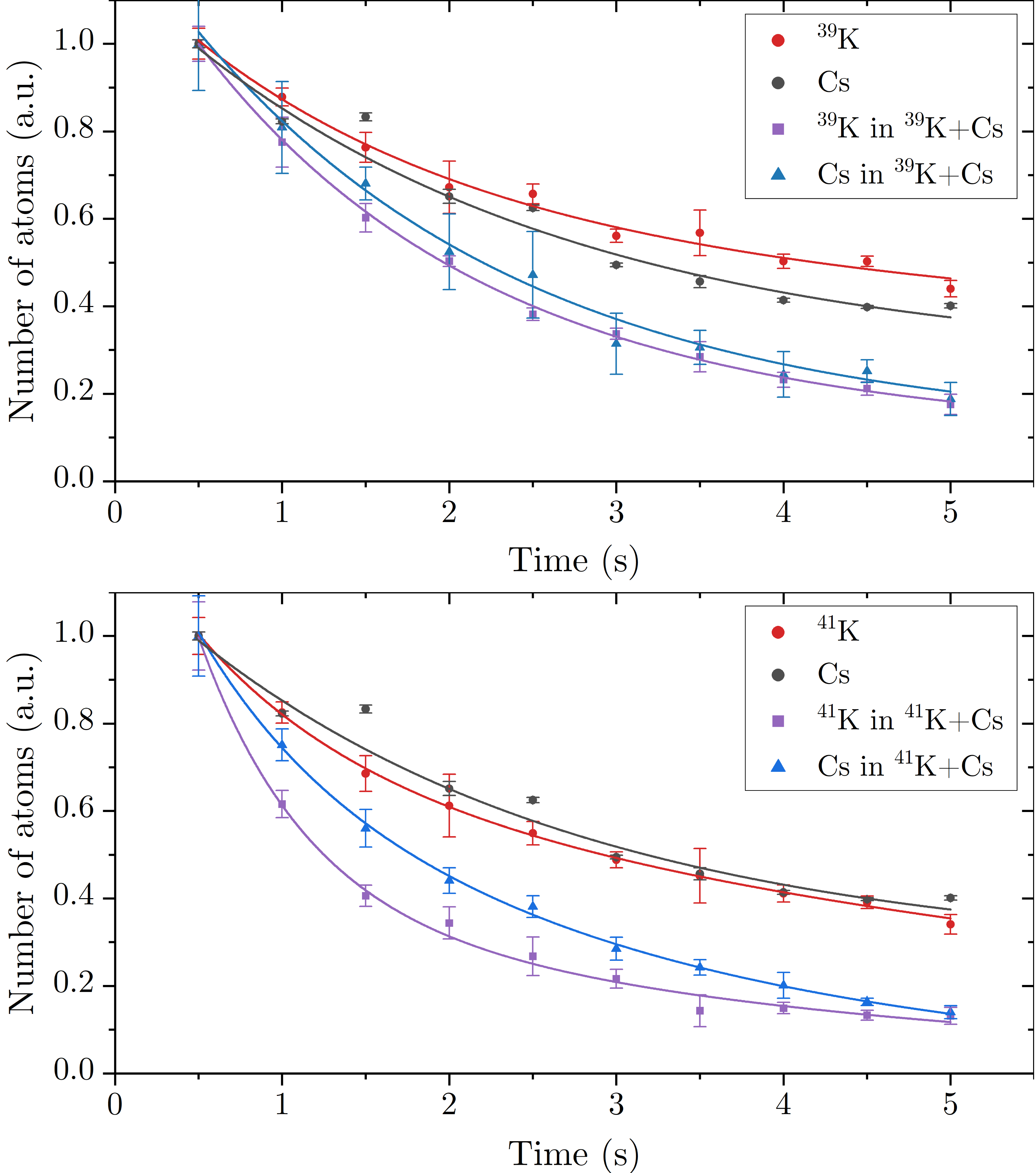} \caption{Normalized atom-number decay in the optical dipole trap for single-species samples and for both components of the K-Cs mixtures. The upper panel shows measurements for the $^{39}$K-Cs mixture and the lower panel for the $^{41}$K-Cs mixture. In each panel, the decay of potassium and cesium in the mixture is compared with the corresponding single-species decay. The solid curves are phenomenological double-exponential fits used to characterize the fast and slow components of the observed decay. We do not assign fitted time scales to individual microscopic loss processes. Atom numbers are normalized to
their values at the first measured trapping time.} \label{41KCs_lifetime} 
\end{figure}

Figure~\ref{41KCs_lifetime} compares the decay of the individual species with the decay of both components of the corresponding K-Cs mixtures. For both potassium isotopes, the presence of Cs leads to a clear acceleration of the potassium decay. Conversely, cesium also decays faster in the presence of potassium than in the corresponding single-species sample. The accelerated decay of both constituents therefore demonstrates that the additional loss is associated with the presence of the second species rather than with a species-specific background loss mechanism.

The decay curves are not well described by a single exponential, particularly for the $^{41}$K-Cs mixture. We therefore use a phenomenological double-exponential function,
\begin{equation}
N(t)=N_0+\sum_{i=1}^2 A_i\exp[-(t-t_0)/\tau_i],
\end{equation}
to characterize the evolution of the atom number. Here, $t_0=0.5$~s corresponds to the first measured trapping time. The two exponential components should be regarded as characteristic time scales of the observed decay rather than as independently identified microscopic loss processes.

The difference between the two isotopic mixtures is particularly pronounced for potassium. For $^{41}$K in the mixture, the fit gives a short-term decay with
$\tau_1=0.63(15)$~s and a long-term decay with
$\tau_2=3.83(1.05)$~s, with the short-term relative amplitude
of approximately 62\%. In comparison, single-species $^{41}$K exhibits
$\tau_1=0.94(27)$~s and $\tau_2=6.94(1.32)$~s, with a relative amplitude
of approximately 33\% for the short-term decay. A similar enhancement of the decay is observed for cesium in the $^{41}$K-Cs mixture, for which the fitted time scales are $\tau_1=0.72(34)$~s and $\tau_2=2.67(28)$~s.

For the $^{39}$K-Cs mixture, the additional loss is weaker. The short-term component of the $^{39}$K decay has a characteristic time of approximately 1.7~s, while the long-term component extends beyond the duration of the present measurement and is consequently not well constrained by the fit. The same limitation applies to the long-term component of some of the single-species and cesium reference measurements. We therefore do not assign physical significance to the individual long-term constants in these cases.

The faster initial decay followed by a progressively slower loss is consistent with a density-dependent contribution to the decay. Possible contributions include heteronuclear few-body processes, in particular three-body recombination, while one-body losses due to collisions with the background gas become relatively more important at longer trapping times. A quantitative extraction of the corresponding heteronuclear loss coefficients would, however, require a rate-equation analysis including the time evolution of the densities and temperatures of both species. We therefore restrict the present analysis to a phenomenological characterization of the observed decay dynamics.

The two isotopic mixtures have substantially different low-energy scattering properties. The most recent values available in the literature give $a_{\mathrm{s}}^{39\mathrm{K-Cs}}=-29.2\,a_0$ and $a_{\mathrm{t}}^{39\mathrm{K-Cs}}=77.7\,a_0$, compared with $a_{\mathrm{s}}^{41\mathrm{K-Cs}}=-72.79\,a_0$ and $a_{\mathrm{t}}^{41\mathrm{K-Cs}}=179.06\,a_0$~\cite{PhysRevA.95.022715,beulenkamp2025formation}. Although the underlying singlet and triplet interaction potentials are approximately isotope independent, the change in reduced mass modifies the near-threshold scattering properties and results in substantially different scattering lengths for the two isotopologues. This provides a natural basis for expecting different collisional behavior of $^{39}$K-Cs and $^{41}$K-Cs. The singlet and triplet scattering lengths, however, do not by themselves determine the inelastic loss rates in the specific hyperfine channels used in our experiment.

\section{Conclusions}

In this work, we have demonstrated simultaneous sub-Doppler cooling and optical trapping of $^{39}$K-Cs and $^{41}$K-Cs mixtures in a single experimental platform. By optimizing the sub-Doppler cooling and optical-trap loading procedures, we have achieved efficient transfer of both potassium and cesium into the optical dipole trap, with transfer efficiencies reaching up to 7\%. Both mixtures are prepared with more than $10^6$ atoms per species at temperatures of approximately 10--15~\textmu K, comparable to those obtained in single-species operation. To the best of our knowledge, this constitutes the first realization of a laser-cooled and optically trapped $^{41}$K-Cs mixture.

Using the $^{39}$K-Cs mixture, we perform multichannel Feshbach spectroscopy and observe 14 heteronuclear loss features. Five correspond to resonances reported previously, while nine have, to the best of our knowledge, not been observed experimentally before. Spin-resolved detection based on Stern--Gerlach separation allows the three potassium $F=1$ spin components to be measured independently within the same experimental sequence, reducing the number of experimental cycles required for multichannel resonance searches and providing direct identification of the potassium entrance channel associated with each loss feature without the need for optical pumping.

We additionally characterize the stability of the two isotopic mixtures under comparable experimental conditions. The $^{41}$K-Cs mixture exhibits substantially faster decay than $^{39}$K-Cs, with an additional density-dependent contribution visible in the atom-number evolution. Measurements of both components show that the presence of the second species accelerates the decay of both potassium and cesium. The microscopic origin and order of the corresponding inelastic loss processes cannot be determined from the present measurements and will require a dedicated study.

The cooling approach should also be applicable to the fermionic $^{40}$K-Cs mixture. We do not explore this combination here because the low natural abundance of $^{40}$K in our potassium source limits the atom number to approximately $1\text{-}5\times10^5$ at the MOT stage.

The demonstrated preparation of ultracold $^{41}$K-Cs provides a starting point for systematic Feshbach-resonance and photoassociation spectroscopy of this previously unexplored isotopologue. Together with the additional spectroscopic information obtained for $^{39}$K-Cs, these results extend the experimentally accessible potassium--cesium systems and provide the basis for identifying association pathways toward ultracold ground-state KCs molecules.

\section*{Acknowledgments}
We acknowledge M. Frye and M. Tomza for fruitful discussions and for the initial calculations of the expected resonance positions based on the available interaction potential. This research was funded by the Foundation for Polish Science within the Homing program and the National Science Centre of Poland (grant No. 2016/21/D/ST2/02003, No. 2021/43/B/ST4/03326 and a postdoctoral fellowship for M.S., grant No. DEC-2015/16/S/ST2/00425).

\section*{DATA AVAILABILITY}
The data that support the findings of this article are openly available~\cite{G8KYWM_2026}.

\newpage

\bibliography{bibliography}

\clearpage
\onecolumngrid
\setcounter{figure}{0}
\renewcommand{\thefigure}{S\arabic{figure}}

\setcounter{table}{0}
\renewcommand{\thetable}{S\arabic{table}}

\setcounter{equation}{0}
\renewcommand{\theequation}{S\arabic{equation}}

\setcounter{section}{0}
\renewcommand{\thesection}{S\arabic{section}}
\section*{SUPPLEMENTARY MATERIAL} \vspace{1 cm}
\section{Experimental setup and single species cooling}

\subsection{Vacuum system and magnetic-field coils}

\begin{figure}[ht]
\centering
\includegraphics[width=0.5\columnwidth]{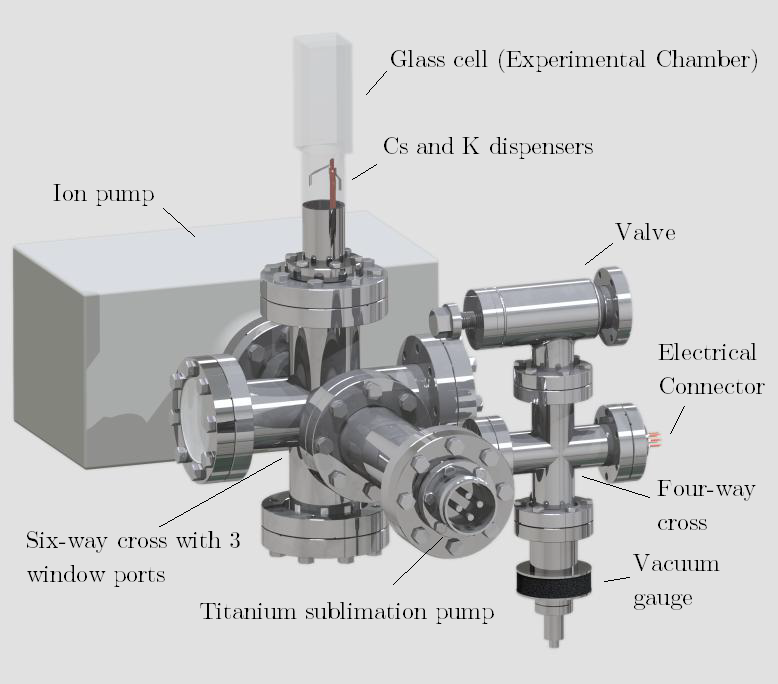}
\caption{Rendering of the experimental chamber.}
\label{fig:chamber_supp}
\end{figure}

The vacuum system is based on a single-chamber design, with atomic dispensers (SAES getters) located 7~cm from the center of the magneto-optical trap, serving as a source of atoms (see Fig.~\ref{fig:chamber_supp}). In the system, two dispensers per species are available. During experiments with cold potassium and cesium mixtures, we typically run a single dispenser per species at 3.2--3.6~A to maintain a satisfactory balance between the vacuum lifetime and the maximum number of trapped atoms. The potassium dispensers contain a natural isotopic composition, consisting of 93.3\% of $^{39}$K, 0.012\% of $^{40}$K and 6.73\% of $^{41}$K~\cite{tiecke2010properties}.

The base pressure, when the dispensers have been off for 1--2 days, reaches 2--4$\times 10^{-11}$~mbar. During measurements, both types of dispensers remain continuously powered, increasing the background pressure and consequently reducing the background-gas-limited trap lifetime. The magnetic field generated by the dispensers at the location of the trap is negligible, as verified by RF spectroscopy of $^{41}$K atoms.

Three pairs of compensation coils are arranged symmetrically around the glass cell to form a cage. A detailed description of the compensation coil system and its driving current source is provided in~\cite{dobosz2021bidirectional}.

Helmholtz and anti-Helmholtz coil pairs are mounted radially near the glass cell, approximately 12~cm from the sample of cold atoms. The coil holders include slits to suppress eddy currents that may arise during fast field switching. The coils are powered by "Delta Elektronika SM45-70D" and "SM70-45D" power supplies, which can deliver up to 45~A and 70~A to the anti-Helmholtz and Helmholtz coils, respectively. This limits the available gradient to about 115.2~G/cm along the coil axis and 57.6~G/cm in transverse directions.

\subsection{Single-species cooling parameters}

The notation for detunings, natural linewidths, and saturation intensities follows that introduced in the main text. The duration of each cooling step, magnetic fields, beam detunings, their intensities, atom numbers, temperatures, and phase-space densities in the case of a single-species operation are summarized in Table~\ref{table_cooling_supp}.

\begin{table*}[]
\resizebox{\textwidth}{!}{
\begin{tabular}{cccccccccc}
\hline
Stage                                                                         & Atom                      & \begin{tabular}[c]{@{}c@{}}$\partial B/  \partial z$\\  (G/cm)\end{tabular} & $\delta_{\mathrm{C}}$ $(\Gamma)$ & $\delta_{\mathrm{R}}$ $(\Gamma)$ & $I_{\mathrm{C}}$ $(I_s)$ & $I_{\mathrm{R}}$ $(I_s)$ & N ($\times 10^6$)    & T ($\mu$K)            & $\rho$                               \\ \midrule
\multirow{3}{*}{\begin{tabular}[c]{@{}c@{}}MOT\\ (10-60~s)\end{tabular}}      & $^{39}$K                  & 6                                                                           & -8                               & -5                               &                          24.3&                          10.1& 250                  & 2500                  & $8.5\times 10^{-10}$                 \\
                                                                              & $^{41}$K                  & 6                                                                           & -6.2                             & -6.2                             & 23                       & 17.5                     & 43                   & 10000                 & $10^{-11}$                           \\
                                                                              & $^{133}$Cs                & 5.2                                                                         & -3                               & -2.5                             &                          13.1&                          2.2& 120                  & 155                   & $4.6\times 10^{-8}$                  \\ \midrule
\multirow{3}{*}{\begin{tabular}[c]{@{}c@{}}CMOT\\ (10-15~ms)\end{tabular}}    & $^{39}$K                  & 55                                                                          & -9                               & -7                               &                          21&                          6.2& 240& 11000                 & $9.7\times 10^{-10}$                 \\
                                                                              
                                                                              & $^{41}$K                  & 35                                                                          & -3                               & -3.5                             & 23                       & 17.5                     & 42                   & -                     & -                                    \\
                                                                              & $^{133}$Cs                & 23                                                                          & -3.5                             & -3.5                             &                          12.8&                          1.8& 100                  & 200                   & $5.2\times 10^{-8}$                  \\ \midrule
\multirow{5}{*}{\begin{tabular}[c]{@{}c@{}}D2D1-CMOT\\ (2.6~ms)\end{tabular}} & $^{39}$K$_{\mathrm{D2}}$  & \multirow{2}{*}{55}                                                         & -8                               & -4                               &                          12.2&                          6& \multirow{2}{*}{210}& \multirow{2}{*}{140}  & \multirow{2}{*}{$3.6\times 10^{-7}$} \\
                                                                              & $^{39}$K$_{\mathrm{D1}}$  &                                                                             & -1                               & +8                               &                          15.6&                          12.8&                      &                       &                                      \\
                                                                              
                                                                              & $^{41}$K$_{\mathrm{D2}}$  & \multirow{2}{*}{35}                                                         & -3                               & -3.5                             & 11.5                     & 9                        & \multirow{2}{*}{40}  & \multirow{2}{*}{1000} & \multirow{2}{*}{$9.7\times 10^{-9}$} \\
                                                                              & $^{41}$K$_{\mathrm{D1}}$  &                                                                             & +4.5                             & +7.5                             & 7.3                      & 7.3                      &                      &                       &                                      \\
                                                                              & $^{133}$Cs                & -                                                                           & -                                & -                                & -                        & -                        & -                    & -                     & -                                    \\ \midrule
\multirow{3}{*}{\begin{tabular}[c]{@{}c@{}}GMC\\ (8-14~ms)\end{tabular}}      & $^{39}$K$_{\mathrm{D1}}$  & -                                                                           & +4 (TP)                               & +13.2 (TP)                            &                          15.6 $\to$ 2.2&                          11.7 $\to$ 3& 200                  & 13                    & $1.2\times 10^{-5}$                  \\
                                                                             
                                                                              & $^{41}$K$_{\mathrm{D1}}$  & -                                                                           & +5 (TP)                          & +10 (TP)                         & 14.6 $\to$ 4.6           & 13.6 $\to$ 4.2           & 38                   & 11                    & $1.7\times 10^{-5}$                  \\
                                                                              & $^{133}$Cs                & -                                                                           & -42.3                            & -38.6                            & 4.5                      & 1.5                      & 75& 23                    & $2.5\times 10^{-5}$                  \\ \midrule
\multirow{3}{*}{\begin{tabular}[c]{@{}c@{}}TO-GMC (20-60 ms)\end{tabular}}& $^{39}$K$_{\mathrm{D1}}$  & -                                                                           & +4 (TP)                               & +13.2 (TP)                           &                          1.4&                          0.3& 150& 16                    & $8\times 10^{-6}$                    \\
                                                                              
                                                                              & $^{41}$K$_{\mathrm{D1}}$  & -                                                                           & +5 (TP)                          & +10 (TP)                         &                          2.2&                          0.7& 35                   & -                     & -                                    \\
                                                                              & $^{133}$Cs                & -                                                                           & -42.3                            & -38.6                            & 3.3& 0.35& 36                   & 24                    & $5.5\times 10^{-8}$                  \\ \midrule
\multirow{3}{*}{Dipole trap}                                                  & $^{39}$K                  & -                                                                           & -                                & -                                & -                        & -                        & 3.6                  & 7.5                   & $1.0\times 10^{-4}$                  \\
                                                                             
                                                                              & $^{41}$K                  & -                                                                           & -                                & -                                & -                        & -                        & 2                    & 7.5& $3.2\times 10^{-5}$                  \\
                                                                              & $^{133}$Cs                & -                                                                           & -                                & -                                & -                        & -                        & 5.1& 12& $3.0\times 10^{-5}$\\ \midrule
\end{tabular}
}
\caption{Main cooling stages in the single-species experimental procedure. The first column lists the cooling stages, using abbreviations defined in the main text. The corresponding durations for different atomic species are given in parentheses. The subsequent columns list magnetic field gradient $\partial B/ \partial z$ along gravity, detunings from cooling $\delta_{\mathrm{C}}$ and repumping $\delta_{\mathrm{R}}$ transitions, intensities of cooling $I_{\mathrm{C}}$ and repumping $I_{\mathrm{R}}$ beams, atom number $N$, averaged typical temperature $T$ and phase space density $\rho$. For the stages in which the two-photon condition is essential, we emphasize it by (TP) in parentheses next to the frequency. The cooling and repumping transitions, natural linewidth $\Gamma$, and saturation intensity $I_{\mathrm{s}}$ are species-dependent, as explained in the main text. Temperature and phase-space density values for $^{41}$K are omitted for stages in which the high temperature or distortion of the cloud shape made the measurements unreliable. The cloud densities entering the phase-space-density calculation are obtained from Gaussian fits to the measured density distributions.}
\label{table_cooling_supp}
\end{table*}

\subsection{Additional details of potassium cooling}

The number of atoms in the magneto-optical trap typically saturates after 10--60~s, depending on the dispenser current and the abundance of the isotope in the source. Although a higher dispenser current increases the atom number and shortens the MOT loading time, it significantly reduces the lifetime of the conservative traps due to the increased background pressure in the experimental chamber. These effects were discussed in detail in our previous work~\cite{bochenski2024sub}.

At the end of the compression stage, we use mechanical shutters to block D2-line light and introduce the D1-line cooling and repumping beams. The switching process takes 2.6~ms, including up to 1.5~ms of timing jitter and 1.1~ms of shutter switching time.

After the D2D1-CMOT stage, the magnetic field is turned off in 0.3~ms. At the same time, the output acousto-optic modulators are switched to an idle frequency within 0.1~ms, at which no light is sent to the experiment, while the modulators remain operating. This approach reduces the thermal effects that could arise from inactivity. During the switching interval, the relatively high cloud temperature of 2--10~mK leads to free expansion and a measurable reduction in density.

The 1064~nm dipole-trap beam has a waist of 230$\times$240~\textmu m in the focus, and each arm has a power of 32~W. Because optical losses on the uncoated cell surfaces make it difficult to infer the depth of the trap directly from the incident power, we characterize the trap by parametric heating~\cite{photonics9070442}. Using this method for potassium atoms, we obtained trap frequencies of $\omega_x,\omega_y,\omega_z = 2\pi \times [60,~135,~148]$~Hz.

For TO-GMC, the frequencies of the cooling and repumping beams are kept unchanged, and only their intensities and the duration of the stage are varied. Our earlier work showed that properly optimized GMC in our system can hold the cloud for more than 1~s, but we choose to limit the holding time to 100~ms to mitigate losses from collisions with background gas~\cite{dobosz2021bidirectional}. The maximum transfer efficiency is obtained after $t=50$--60~ms. The corresponding optimization measurements are shown in Fig.~1 of the main text.

The last stage of the sample preparation is spin polarization in the selected magnetic sub-level. In the case of bosonic potassium, we turn on the uniform magnetic field $B=2.5$~G, and switch on two circularly polarized beams resonant with $^2S_{1/2}F=1 \to {^2P_{1/2}} F=1$, $^2S_{1/2} F=2 \to {^2P_{1/2}} F=2$ transitions. The polarization of beams and the direction of the magnetic field are set in such a way as to drive only $\sigma+$, or $\sigma-$ transitions. After 0.5~ms, we obtain a sample polarized in the $^2S_{1/2}F=2,~m_f=2$ and $^2S_{1/2}F=1,~m_f=1$ states. We turn off these two beams and introduce one that propagates in the opposite direction. This beam is tuned to the $^2S_{1/2} F=2 \to {^2P_{1/2}} F=2$ transition and has a linear polarization, exciting only the $\pi$ transition.

\subsection{Cesium laser system}

The system consists of two Toptica TA-Pro MOPA systems, stabilized with a CoSy (compact spectroscopy) module and Digilock 110. The repumping laser is stabilized to the $F=3 \rightarrow F'=3$ transition, while the cooling laser has an acousto-optic modulator in the stabilization path, used to allow stabilization 69~MHz above the $F=4 \rightarrow F'=4$ transition frequency. The lasers, unlike in the case of potassium (see Ref.~\cite{bochenski2024magnetic}), are stabilized to the lock-in detected transitions obtained by modulation of the diode current at 41.12~kHz and 45.96~kHz for the repumping and cooling lasers, respectively. These modulations result in a broadening of the effective spectral width to about 4~MHz. A more detailed discussion of this effect can be found in work~\cite{Koray}.

The light from the optical amplifiers passes through dedicated mechanical shutters, followed by single-mode polarization-maintaining fibers used to propagate light to a second optical table, where fine-tuning of the frequency and free-space propagation to the experimental system takes place.

\begin{figure*}[ht]
\centering
\includegraphics[width=\textwidth]{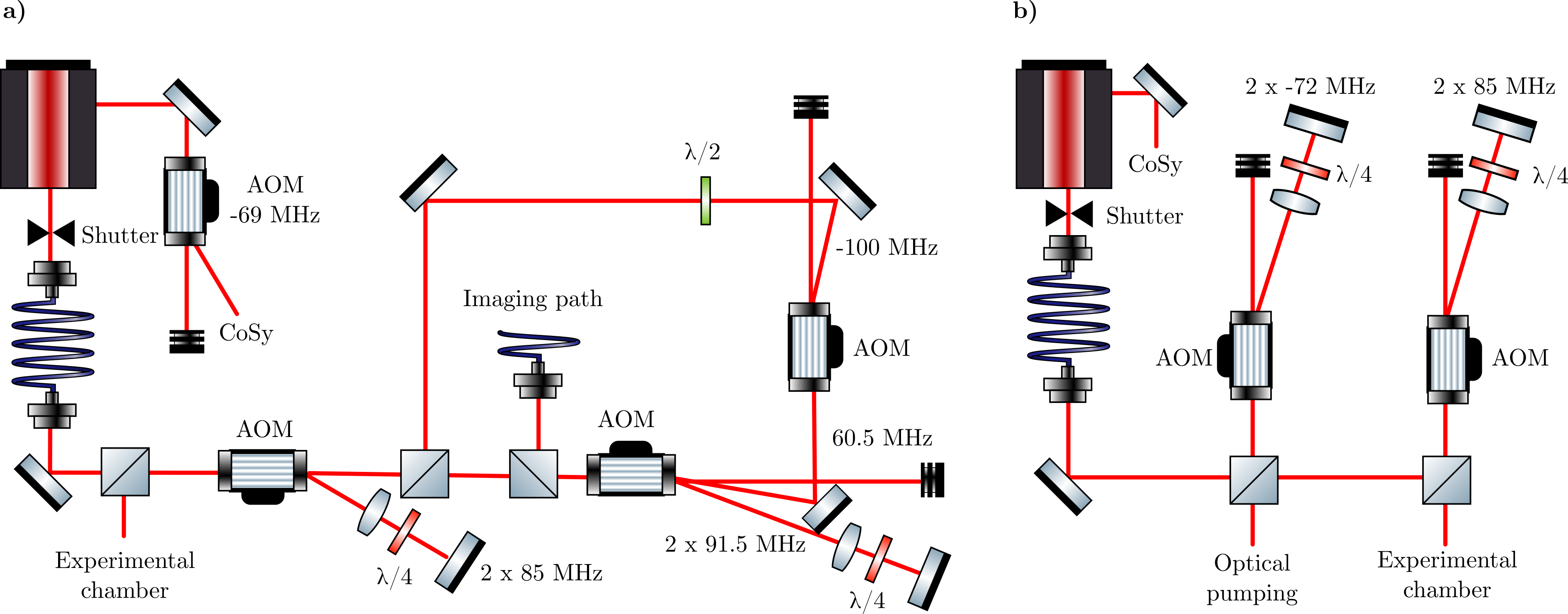}
\caption{Laser system for cooling cesium atoms, presented in two parts. a) Distribution of light stabilized 69 MHz over $F=4 \rightarrow F'=4$ transition in Compact Spectroscopy (CoSy) module. Light is distributed and frequency-tuned using acousto-optical modulators, where the default deflection frequency is marked in the figure, with the multiplier indicating a double-pass configuration. b) Distribution of light stabilized to $F=3 \rightarrow F'=3$ transition.}
\label{Cesium_laser_system_supp}
\end{figure*}

The repumping beam is divided into two paths, as presented in Fig.~\ref{Cesium_laser_system_supp}b). Each path contains an acousto-optic modulator (AOM) in a double-pass configuration. One of the modulators operates around its central frequency of 80~MHz, and its main purpose is to tune repumping light for various stages of the experiment. This modulator is also used as a fast switch (sub-$\mu$s).

On the second path, there is an AOM operating at --72~MHz, so the beam is blue-detuned by $\Gamma_{\mathrm{Cs}}$ from the $F=3 \rightarrow F'=2$ transition. After the double pass, this beam overlaps with one of the beams dedicated to potassium experiments and is sent to the vacuum system via an optical fiber. A linear polarizer and an achromatic quarter-wave plate are placed at the output of the fiber, in such a way as to obtain an optical path whose purpose is to optically pump atoms between magnetic sublevels.

The path dedicated to fine-tuning the frequency of light stabilized 69~MHz over $F=4 \to F'=4$ involves more components, as it must accommodate both cooling and probing purposes. The beam from the optical fiber is deflected by an AOM operating at 85~MHz in a double-pass configuration. In this way, the light is tuned 2.5$\Gamma_{\mathrm{Cs}}$ below the cooling transition $F=4 \rightarrow F'=5$.

With the modulator turned off, the beam passes to a second AOM set to +91.5~MHz in a double-pass configuration. In this configuration, the beam is resonant with the $F=4 \rightarrow F'=5$ transition and is coupled into an independent optical fiber, enabling absorption imaging of the atoms.

If the modulator operates instead at +60.5~MHz, the beam is deflected in the single-pass configuration. The deflected beam then passes through another AOM operating at -100~MHz, becoming blue-detuned by 6$\Gamma_{\mathrm{Cs}}$ relative to the $F=4 \rightarrow F'=4$ transition. The beam is then superimposed with the MOT cooling beam and distributed in the system. This beam is used only during the optical molasses cooling stage.

\subsection{Additional details of cesium cooling}

The parameters of the MOT, CMOT, GMC, and TO-GMC stages are given in Table~\ref{table_cooling_supp}. During CMOT compression, the magnetic gradient increases to 23~G/cm in 2~ms. When the field reaches the desired value, the compression process takes 8~ms.

At the beginning of gray-molasses cooling, we use all the available cooling-light power. After 8~ms, the beam intensities change to the values required for TO-GMC. In the cesium case, thermal effects in the acousto-optic modulator cause a small drift in the propagation direction of the cooling beam. Since the beam travels over a distance exceeding 2~m before reaching the atoms, this angular drift results in a noticeable displacement of the beam position during prolonged cooling. If cooling takes longer than 60~ms, the GMC cooling beams become misaligned and the transfer efficiency decreases. The corresponding optimization measurements are shown in Fig.~1 of the main text.

Finally, in the optical-pumping stage, we applied a magnetic field of 2.5~G and introduced a laser beam blue-detuned by 1.5$\Gamma_{\mathrm{Cs}}$ from the $F=3\to F'=2$ transition, with circular polarization. The magnetic-field direction is chosen such that the beam predominantly drives $\sigma^+$ transitions, while retaining a non-negligible $\pi$-polarized component. During optical pumping, a beam resonant with the $F=4 \to F'=5$ transition is constantly turned on to excite atoms that could decay to the $F=4$ state from off-resonant excitations.


\end{document}